# Improving Task Instructions for Data Annotators

How Clear Rules and Higher Pay Increase Performance in Data Annotation in the AI Economy

Johann Laux♦♥✉, Fabian Stephany♦♥♠, Alice Liefgreen♣

♦ Oxford Internet Institute, University of Oxford, UK,
♥ Humboldt Institute for Internet and Society, Berlin,
♠ Bruegel, Brussels, Belgium,
♣ University College London
✉ [johann.laux@oii.ox.ac.uk](mailto:johann.laux@oii.ox.ac.uk)

## ABSTRACT

*The global surge in AI applications is transforming industries, leading to displacement and complementation of existing jobs, while also giving rise to new employment opportunities. Data annotation, encompassing the labelling of images or annotating of texts by human workers, crucially influences the quality of a dataset directly influences the quality of AI models trained on it. This paper delves into the economics of data annotation, with a specific focus on the impact of task instruction design (that is, the choice between rules and standards as theorised in law and economics) and monetary incentives on data quality and costs. An experimental study involving 307 data annotators examines six groups with varying task instructions (norms) and monetary incentives. Results reveal that annotators provided with clear rules exhibit higher accuracy rates, outperforming those with vague standards by 14%. Similarly, annotators receiving an additional monetary incentive perform significantly better, with the highest accuracy rate recorded in the group working with both clear rules and incentives (87.5% accuracy). In addition, our results show that rules are perceived as being more helpful by annotators than standards and reduce annotators' difficulty in annotating images. These empirical findings underscore the double benefit of rule-based instructions on both data quality and worker wellbeing. Our research design allows us to reveal that, in our study, rules are more cost-efficient in increasing accuracy than monetary incentives. The paper contributes experimental insights to discussions on the economical, ethical, and legal considerations of AI technologies. Addressing policymakers and practitioners, we emphasise the need for a balanced approach in optimising data annotation processes for efficient and ethical AI development and usage.*

I. INTRODUCTION

Over the past decade, Artificial Intelligence (AI) systems have significantly improved their performance and are increasingly being used by the general public (Maslej et al., 2024). This surge in AI uptake has led to a rising demand for AI-related human labour. The high economic value of AI skills (Stephany & Teutloff, 2024) is reflected in higher pay of up to 23% for workers with AI skills within the same occupation (Gonzalez Ehlinger & Stephany, 2023). Besides well-paid engineering work, human labour is also needed to prepare the data used to design and develop AI systems. These "AI data workers" collect, curate, annotate, and evaluate datasets and AI model outputs (Muldoon, Cant, et al., 2024, pp. 1–2).

This paper is concerned with one particular task in AI data work – namely, annotation. Annotating data sets is essential to training, testing, and validating AI and machine learning models (Patra et al., 2023). It is the biggest and most time-costly part of data preparation in the AI lifecycle (Muldoon, Cant, et al., 2024, p. 10). Human annotation, including labelling images and tagging texts, directly impacts the efficiency and accuracy of AI models (Rädsch et al., 2023; Gupta et al., 2021; Shad et al., 2021; Willemink et al., 2020), and it can be as crucial to the performance of an AI model as computational power (Eisenmann et al., 2023). *ImageNet* is arguably the most prominent example for the pivotal role of a human curated dataset for AI applications. It contains more than 14 million hand-labelled images of objects (ImageNet, 2020). Since its publication in 2009, *ImageNet* has led to a boom in deep learning for machine learning (Ruder, 2018) and has been leveraged by big tech companies to transfer learning and develop AI solutions for tasks such as object detection (He et al., 2017) and action recognition (Carreira & Zisserman, 2017). Data annotation plays a pivotal role in mitigating biases and ensuring fairness in predictions (Chen et al., 2023), but it can also introduce biases and lead to discriminatory outcomes (Denton, Hanna, et al., 2021; Paullada et al., 2020; Yang et al., 2020). Labelling errors in commonly used test data sets such as *ImageNet* can have adverse effects for the selection of AI models, as AI practitioners often choose which model to deploy based on accuracy benchmarked against the test data set (Northcutt et al., 2021).

The global data annotation market is currently valued at 1.3 billion USD and is expected to reach 14 billion USD by the end of 2035 (*Data Annotation Tools Market*, 2022). Human-led data annotation has thus become an economy of global relevance and a significant economic factor in the production of AI applications. To reduce costs, methods are being developed to minimize the required amount of human-labelled data or to automate data annotation (Wang et al., 2023). Firms also outsource annotation work (Tubaro, 2021), both to outsourcing companies and online labour platforms (Muldoon, Cant, et al., 2024; Schmidt, 2022; Miceli & Posada, 2022; Tubaro et al., 2020; Ørting et al., 2020). The outsourced labour often goes to countries in the Global South, where wages are lower and labour regulations are less defined (Le Ludec et al., 2023; Braesemann et al., 2022; Graham & Anwar, 2019). This development has led to debates about fair compensation and work conditions for data annotators (Muldoon, Graham, et al., 2024; Wong, 2023; Gray & Suri, 2019; Wood et al., 2019). In 2023, *Times* reported that OpenAI had outsourced the task of labelling texts with descriptions of violence, hate speech, and sexual abuse to train an AI-driven safety system for its large language model ChatGPT to Kenyan workers who earned less than $2 per hour (Perrigo, 2023).

The discussion surrounding the working conditions of AI data workers is not merely about improving the lives of millions of people around the world (Kässi et al., 2021). It also has a direct impact on the quality of AI applications that are built with data prepared by humans (Chen et al., 2023; Rädsch et al., 2023; Shemtob et al., 2023; Wang et al., 2023; Yuan et al., 2023; Denton, Díaz, et al., 2021; Gehman et al., 2020; Ørting et al., 2020). This paper therefore asks a foundational economic question about AI data work and worker wellbeing: *Do AI developers have an incentive to invest in improving the working conditions of data annotators?* We stipulate that this incentive would lie in higher accuracy in annotation outcomes, leading to higher-quality data sets. We are thus interested in the influence of work conditions on annotation accuracy.

We pursue our research question along two aspects of data annotators' work conditions: the formulation of their task instructions and their pay. Previous research has shown that how task instructions are formulated on online labour platforms impacts the quality of work outcomes (Gillier et al., 2018). According to a recent survey, data annotators in biomedical image labelling stated that the greatest cause of problems in their daily work were unclear task instructions (Rädsch et al., 2023, p. 275). We investigate the formulation of task instructions by drawing on a distinction common in the field of law and economics: the difference between formulating norms as rules or standards (Kahneman et al., 2021; Korobkin, 2000). This choice allows us to vary the amount of task information provided to annotators from a cost-based vantage point. Rules are more costly than standards for the norm-setter, as they require more *a priori* collection of information (See Section II for further information on rules vs. standards). We further postulate that raising workers' pay level can increase their motivation and, hence, their effort. Psychological research has found a positive link between financial incentives and workers' performance (Kim et al., 2022). This is not to say that financial incentives are the only way to raise workers' motivation.

Previous research on online labour platforms has found that the perceived "meaningfulness" of a data annotation task can be influenced by the framing of the task, i.e., whether or not the description of the task appeals to its meaningfulness (Chandler & Kapelner, 2013). However, these findings do not exclude a positive effect on workers' motivation through financial incentives.

To investigate the influence of task formulation and financial incentives on the accuracy of data annotation outcomes, we conducted an experimental study involving 307 data annotators (referred to as annotators henceforth) engaged in an image annotation task. In this study we randomly allocated annotators to one of three different groups provided with either a) rules, b) incomplete rules, or c) standards on how to annotate an image. In addition, annotators in each group received either 1) a regular base-rate pay of £4.50 or 2) an additional monetary incentive to reward higher accuracy of a single payment of £3.

Our findings clearly indicate an association between annotators' performance (measured in accuracy) and investment in their work conditions. Annotators working with rules perform 14% better in terms of accuracy than annotators working with standards. Annotators receiving an additional accuracy-based monetary incentive, on the other hand, show a 3% increase in accuracy. In combination, annotators working under clear rules and with a monetary incentive have the highest accuracy rate in the entire sample (86.7%), which is comparable to today's highest-scoring AI models when undertaking a comparable classification task (Deitz, 2023). Our findings show that information-rich task instructions given via rules, when combined with accuracy-based monetary incentives, are an investment in the economics of AI data work. For AI developers, it pays off to invest in data annotators' work conditions as they receive more accurately annotated data for their AI systems and, hence, are better-positioned to produce higher-quality AI systems. Annotators also perceived rules to be more helpful than standards (a 30% increase) and reported fewer images they had trouble with labelling when working under rules than under standards (a reduction of 22%). That said, while rules and financial incentives improve annotation outcomes, they cost time. The best-performing group – annotators under rules and with monetary incentives – was also the group that needed the most time to complete the task (3.4 seconds to annotate an image, on average). This was 32% longer than the fastest group of annotators, i.e., those working under standards and with no monetary incentive provided. In as much as time factors into labour costs of data annotation, we found rules to be more labour-cost efficient than the monetary incentive in raising accuracy.

To our knowledge, these findings are the first experimental evidence on how conditions of task instructions combined with monetary incentives influence the quality and costs of data annotation. Prior work addressed either the influence of task instructions (Rädsch et al., 2023) *or* monetary compensation (Litman et al., 2015) on data annotation quality. We are also not aware of any other study applying the concept of rules and standards in a data annotation experiment. We believe that our work delivers a relevant contribution to understanding potential quality gains in the market of data annotation, which could ultimately lead to a more efficient, fair, and safe development of AI.

The next section introduces the theoretical background of the economics of data annotation. Section III describes the methods of our experiment and Section IV the results. Lastly, Section V concludes on how our work contributes to the economics of data annotation and the debate on working conditions for AI data workers.

## II. Theoretical Background and Hypotheses

**Rules and Standards in (Behavioural) Law and Economics**

In formulating our task instruction conditions, we applied a distinction common in law and economics, namely that between rules and standards. Its original concern lies with lawmakers who, when enacting laws, must not only agree on the normative substance of a law but also on its form. Scholars of law and economics commonly understand the choice of form as one between rules and standards ((Clermont, 2020; Korobkin, 2000); for a critique, see: (Schauer, 1991; Schlag, 1985)). Rules bind a decisionmaker to a determinate legal result that follows from the presence of one or more triggering facts. Standards give the decisionmaker more discretion. They require the application of a background principle or policy to a particularised set of facts to reach a legal conclusion (drawing on the definitions in: (Korobkin, 2000; Sullivan, 1992)). A frequently used example is that of a speed limit. It may be formulated as a rule that prohibits driving faster than 55 miles per hour. Or it may be phrased as a standard that prohibits driving with excessive speed (Kaplow, 1992). In the latter case, the decisionmaker needs to first determine what amounts to "excessive speed" before applying the standard to a particular driving behaviour. Rules provide increased clarity and predictability in their application due to their upfront specificity, whereas standards are more vague and, hence, more adaptable to the specific context of a given case (Casey & Niblett, 2017). Rules establish their normative content in advance, while standards determine theirs, at least partially, after the fact (Casey & Niblett, 2017).

Scholars of law and economics understand the choice between rules and standards as one of allocating costs between law makers, law appliers, the subjects of law, and society through social costs ((Korobkin, 2000);

see also: (Ehrlich & Posner, 1974; Kaplow, 1992)). Different types of costs arise in different ways for rules and standards. Broadly speaking, rules create higher up-front decision costs for the lawmaker compared to standards (Casey & Niblett, 2017). Error costs arise when the law is over- or underinclusive with respect to the behaviour it allows ((Casey & Niblett, 2017); on the problem of regulatory precision, see: (Diver, 1983)). These costs can be reduced by obtaining more and better information about the governed behaviour, with a reduction of error costs being associated with greater upfront decision costs on the lawmaker (Casey & Niblett, 2017). In classic law-and-economics analysis, rules tend to be advantageous when the behaviour they govern is frequent and homogenous (i.e., has common characteristics): the rule only needs to be formulated once and thus benefits from economies of scale (Kaplow & Shavell, 2002); see also: (Casey & Niblett, 2017). *Vice versa*, for infrequent and heterogenous behaviour, standards may be preferrable, as the costs of providing specific details for all possible future scenarios can be deferred until the adjudication stage ((Kaplow & Shavell, 2002); see also: (Casey & Niblett, 2017)). Lastly, standards are said to have higher uncertainty costs for their subjects than rules, as it is harder for them to assess whether their behaviour will comply with the law or not (Casey & Niblett, 2017).

Behavioural considerations may alter the cost assessments of the classical model (Jolls et al., 1998). While both law makers and law appliers can suffer from bias, standards may be worse affected. For example, it may be more difficult to address problems like hindsight bias for *ex post* adjudication. Standards may thus not necessarily have lower error costs when compared to rules (Casey & Niblett, 2017).

In practice, the choice between rules and standards frequently falls on a continuum between the extremes of pure rules and pure standards (Korobkin, 2000; Schauer, 2003, 2005). Moreover, some argue that when applied by decision makers, rules and standards converge; the adaptive behaviour of law appliers pushes standards towards rules and rules towards standards (Schauer, 2003).

**Rules and Standards in Data Annotation**

Previous work has examined the effects of varying annotation instructions on crowdsourced data annotators' performance in natural language processing tasks (Bragg et al., 2018; Chang et al., 2017; Hossfeld et al., 2014; K. Chaithanya Manam et al., 2019; Manam & Quinn, 2018; Ning et al., 2020; Tokarchuk et al., 2012) and, more recently, biomedical image analysis (Rädsch et al., 2023). We are not aware of a prior experimental study which utilises the framework of rules and standards to examine the problem of efficiently designing task instructions for data annotators. Drawing on the choice between rules and standards allows us to apply an economic lens to the content of data annotation instructions and the costs task formulation creates for an organisation. In applying the rules-and-standards framework, we thus treat the task instructions as similar to norms (albeit not 'laws') under which the AI data workers operate and the formulation of task instructions as similar to the choice of form for norms.

Current practices in content moderation support the plausibility of applying rules and standards to data annotation. Although not part of the AI production chain *per se* (Muldoon, Cant, et al., 2024, p. 2), content moderation is similar to data annotation in so far as it is a process of human review and classification of digital content. Social media platforms, for example, aim to filter out illegal, violent, hateful, abusive, and otherwise offensive content to protect their users and their public image (Gillespie, 2018; Roberts, 2019). While platforms increasingly rely on machine learning for content moderation to keep up with the growth in user-generated content (Gillespie, 2020; Vincent, 2020), automation has thus far not been able to fully eliminate the need for human involvement (Barker & Murphy, 2020; Magalhães & Katzenbach, 2020; Murphy & Murgia, 2019). Today, human moderators still review content that has been flagged by algorithms and classify it as desirable or undesirable (Killeen, 2022).

Anecdotal evidence exists of content moderation at Facebook and its strategic use of both rules and standards. Facebook's Community Standards determine which content is and which content is not allowed to be shared by users on its platform (Meta, n.d.). Whether an image posted on Facebook counts as an instance of, for example, "glorifying violence" must be interpreted by a human (notwithstanding increasing automation mentioned above). The higher the number of content moderators, the higher the possible variation in judgments about violations of the Community Standards ((Kahneman et al., 2021); drawing on: (Marantz, 2020)). However, Facebook has also issued non-public Implementation Standards. By providing explicit directions for handling graphic images, they constrain the moderators' discretion in determining how to address such content ((Kahneman et al., 2021); drawing on: (Marantz, 2020)). Through our law and economics lens, the Community Standards are standards in the sense that their vagueness requires content moderators' discretion to be implemented. The Implementation Standards qualify as rules as they explicitly instruct content moderators about what to do with graphic images. It appears that by formulating rules for implementation, Facebook chose to internalise decision costs up-front to instruct its content moderators.

As mentioned, the classic law-and-economics literature assumes that rules become increasingly cost-advantageous as the conduct they govern becomes more frequent (Kaplow, 1992). In the adjudication of legal cases, it is commonly held that rules save time for adjudicators (Sunstein, 1995). For data annotators, this may be different. Data annotation requires repeated decisions to be made often within seconds, for example, about whether an image depicts violence or not. A complex rule in this context will involve a corresponding increase in the number of checkboxes that annotators must navigate each time they label an image. When applying a standard, however, annotators may intuitively translate the standard into a more straightforward internalised rule as the task progresses, for example, that every image showing a bleeding human is a depiction of violence.[1]. In other words, it may take annotators longer to label an image under rules than under standards. As the size of a dataset increases, the average time spent per annotated data unit (i.e., cost per unit) becomes more relevant. As annotators progress in their task, they may become more efficient in applying rules, thus possibly approaching the time per data unit they would have spent when working under standards (see Figure A2). In data annotation, labour costs (per data unit) are therefore an important factor in the cost calculus of implementing rules versus standards.

**Monetary Incentives**

Whether and how monetary incentives raise workers' motivation and thus improve their performance has been subject to long-standing controversy in psychological research (see the references in (Kim et al., 2022)). Reinforcement theory (Skinner, 1953) and expectancy theory (Vroom, 1964) assume a positive effect of financial incentives on motivation and performance (Kim et al., 2022). These theories also suggest that extrinsic outcomes such as higher pay and intrinsic outcomes such as the work being interesting combine in an additive manner and have a positive effect on workers' motivation and performance ((Kim et al., 2022); drawing on: (Calder & Staw, 1975; Hamner & Foster, 1975; Hendijani et al., 2016)). Self-determination theory (Deci & Ryan, 1985; Gagné & Deci, 2005; Ryan & Deci, 2018) suggests that while (extrinsic) financial rewards can motivate behaviour, they do so at the expense of intrinsic motivation (Deci, 1972). However, the (non-additive) negative effect of extrinsic rewards on intrinsic motivation has only been argued for work in which high intrinsic motivation is crucial for performance; the motivation for boring and non-interesting work is understood to be predominantly extrinsic (Kim et al., 2022). Task interest thus moderates the effect of incentives on performance (Kim et al., 2022). Several meta-analyses have looked at the effects of payment on extrinsic and intrinsic motivation and work performance (Jenkins et al., 1998; Kim et al., 2022; Weibel et al., 2010). They converge on a positive effect of financial incentives for non-interesting work but disagree on the effects for interesting work (Kim et al., 2022).

---

[1] This would support Schauer's convergence thesis of rules and standards (Schauer, 2005).

Examples of work tasks which have been perceived as interesting and non-interesting vary across studies and sector. In the public sector, tasks requiring low policy expertise have been found to be less interesting (Weibel et al., 2010), and lower-level employees have been shown to find their tasks less interesting than managers in the public service ((Buelens & Van Den Broeck, 2007); see also: (Weibel et al., 2010)). This raises the question of whether data annotation should be classified as interesting or non-interesting work.

Labelling images through applying either rules or standards requires very little novel conceptual thinking and creative problem-solving. In as much as standard-applying annotators enjoy more discretion than annotators who are given rules, the former's intrinsic task motivation could be marginally higher than the latter's. However, we assume the intrinsic motivation of annotators in our study to be low in both the rules and standards conditions. Therefore, the debate about the additivity of extrinsic and intrinsic motivation does not affect the task employed in our study – namely that of labelling a set of images of building entrances as either depicting entrances that are barrier-free or non-barrier-free for wheelchair users (see Materials). Since we don't presume that this work is inherently interesting, any rise in extrinsic motivation due to monetary incentives will likely not be counterbalanced by annotators' potential loss of intrinsic motivation. Extrinsically incentivised annotators will likely be motivated to work more diligently and avoid making mistakes if the monetary incentive is performance-based, especially if their performance can be measured in terms of accuracy and their monetary reward is tied to their level of accuracy.

**Our Study**

In sum, the literature suggests that the formulation of task instructions and monetary incentives are influential in determining the performance of workers in data annotation. To test this, we run a study with 307 participants to empirically investigate these dependencies, guided by two research questions. Drawing on the literature reviewed above, there is theoretical support for assuming that the choice between rules and standards, as well as accuracy-based monetary incentives, will influence the performance of data annotators. To account for decision costs and potential unavailability of perfect information when formulating rules, we add a third instruction design, where annotators receive "incomplete rules"; less detailed information than in the rules condition, but more detailed than in the standards condition. Our study is centred on the following questions:

> *RQ1:* What are the effects of different types of instructions – rules, incomplete rules, and standards – on the performance of annotators in an image classification task?
>
> *RQ2:* What are the effects of an accuracy-based monetary incentive on the performance of annotators in an image classification task?

In our study, 307 annotators are recruited from an online labour platform to participate in an A/B test. They are tasked with classifying images under three distinct instruction conditions (rules, incomplete rules, and standards), and are subjected to two payment structures: a regular base-rate pay of £4.50 and a variation with an additional monetary incentive – a single payment of £3. To comprehensively assess the performance in image classification tasks, we collect information on annotators' educational and occupational backgrounds, as well as their experience levels. This allows us to investigate the influence of these factors on performance and discern how they interact with varying instruction types and the monetary incentive.

We test annotators' performance along three metrics. First, we test for accuracy, i.e., the proportion of images annotators labelled correctly. Here, we hypothesise the following:

> *Hypothesis 1a:* We expect accuracy to be higher among annotators working under rules.
>
> *Hypothesis 1b:* We expect accuracy to be higher among annotators working with a monetary incentive.

Secondly, we survey annotators' perceptions of the helpfulness of task instructions. Here, we use two different metrics. We survey the helpfulness of instructions as perceived by annotators, measured on a five-point Likert scale ranging from "hindered a lot" to "helped a lot". We assume that rules increase perceived helpfulness. We further survey task difficulty as perceived by annotators, measured by the number of images that annotators reported having trouble classifying. We assume that rules reduce perceived difficulty. We use these two different metrics because performing an annotation task requires both understanding the instructions and applying them to a data set. Our two metrics survey both steps. Accordingly, we hypothesise the following:

> *Hypothesis 2a:* We expect perceived helpfulness of instructions to be higher for annotators working under rules.

*Hypothesis 2b:* We expect perceived difficulty to be lower for annotators working under rules.

III. METHOD

**Participants**

A total of 307 participants completed the study, after being recruited from the platform Prolific Academic (prolific.co). Participants' ages range from 18 to 65 years, with a mean age of 29.92 years (SD = 9.72). The gender distribution is as follows: 168 male, 135 female, 1 other, and 3 non-binary participants. 83% of the participants have some form of university education (college or higher) and 26% of them have prior experience in data annotation. Table A1 in the Appendix presents information on participants' gender, age, ethnicity, highest level of education, and employment status. Table A1 lists the demographic survey questionnaire participants were asked to answer.

**Design**

We conduct a 2 x 3 (6 conditions) between-subjects design to examine the influence of task instructions and monetary incentives on participants' data annotation performance. Participants are tasked to annotate a dataset of 100 images of building entrances as either barrier-free or non-barrier-free for a person in a wheelchair. The study incorporates two between-subjects factors: "instructions" with three levels (rules, incomplete rules, standards) and "incentive" with two levels (present, absent).

The first between-subjects factor, "instructions", randomly allocates participants to one of three conditions: "rules", "incomplete rules", or "standards". These conditions differ based on the task instructions provided to the participants during the task (see Materials section). The "rules" and "standards" conditions aim to resemble ideal-type rules and standards as characterised in the law and economics literature. The "incomplete rules" condition accounts for the fact that the *ex-ante* determination of the content of rules requires information that may not always be fully available and/or is costly to obtain (hence, introducing either error costs or decisions costs, see discussion above). To create the rules condition, we draw on existing architectural codes for buildings in Germany (Bayerische Architektenkammer, 2019; Bundesfachstelle Barrierefreiheit & Deutsche Rentenversicherung Knappschaft-Bahn-See, 2020; Senatsverwaltung für Stadtentwicklung, Bauen und Wohnen, 2021), the location of the majority of the entrances in the dataset of images that the annotators are asked to work with.

For the second between-subjects factor, "incentive", participants are randomly assigned to either the "incentive present" group or the "incentive absent" group. The "incentive present" group is informed about the possibility of receiving a performance-determined bonus, which could amount to an additional £3 on top of the baseline pay (further details provided in the Procedure section). In contrast, the "incentive absent" group receives the baseline pay of £4.50 without any additional monetary incentive.

To summarise, participants are randomly assigned to one of six between-subjects conditions: Rules-Incentive Present, Rules-Incentive Absent, Incomplete Rules-Incentive Present, Incomplete Rules-Incentive Absent, Standards-Incentive Present, Standards-Incentive Absent. Each participant engages in our data annotation task by viewing and classifying a total of 100 images during the task. Images are kept the same across participants but are randomised in order of appearance.

**Materials**

*Data Annotation Image Classification Task*

The data annotation task was programmed and completed by participants in Gorilla (https://gorilla.sc), an online research tool for experimental psychology. Participants view and labell 100 images of public, mostly commercial building entrances. The choice of images responds to the difficulty of current machine learning models to predict the accessibility of urban spaces to wheelchair users (Deitz, 2023), thus representing an excellent example of an area that still needs human involvement. The 100 images were hand-selected by one of the researchers from an existing open access dataset on accessibility barriers, consisting of 5,538 images of building entrances in public spaces in Central Europe and the UK (Stolberg, 2022). The selection of the 100 images is based on clear visibility of the architectural features of the entrances and variety as regards these features: stairs, steps, ramps, handrails/grab bars. Half of the images had been previously labelled as barrier-free, and the other half as non-barrier-free by one of the researchers, according to the instructions in the "rules" condition (see below). We tested agreement on annotating the image set amongst two researchers (n=20 images) and report an agreement score of 100%. Participants view the images one at a time, in randomised order, and have to classify them as being "barrier-free" or "non-barrier-free". There is no time limit for each trial. Prior to beginning the task, participants are provided with task instructions (rules, incomplete rules, or standards, depending on the condition), which define what comprises a barrier-free entrance. These instructions are visible during each trial (see Figure 1 for an example trial).

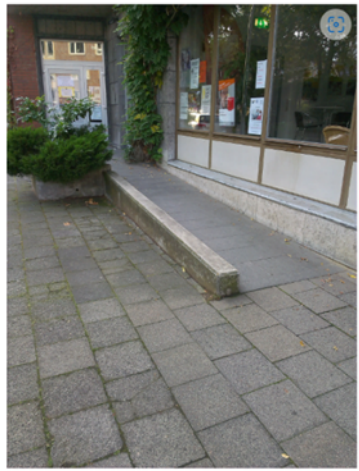

**Figure 1.** Example trial: image of an entrance together with the rule condition as shown to participants.

*Task Instructions*

Participants in the "rules" condition are given the following instructions, before and during the task:

> "(i) Barrier-free entrances must be step-less, (ii) An entrance with a doorstep or curb with a maximum height of 2cm (0,79 inches) is still considered barrier-free. Judging the height by eye will suffice and should proceed with the following question in mind: "Does the picture show a stair (not barrier-free) or a mere doorstep or curb (barrier-free)?", (iii) Ramps must be easy to use and safe for traffic to be barrier-free. Ramps must feature a handrail."

Participants in the "incomplete rules" condition are given the following instructions, before and during the task:

> "(i) Barrier-free entrances must be step-less. If technically necessary, an entrance with a small doorstep or curb is still considered barrier-free, (ii) Ramps must be easy to use and safe for traffic to be barrier-free."

Finally, participants in the "standards" condition are given the following instructions, before and during the task:

> "To be barrier-free, entrances must be accessible for wheelchair users without help by another person."

*Monetary Incentive*

Participants in the "incentive" conditions, are given the following information before beginning the task:

> "In addition to the base-rate payment at a rate of £4.50 per hour, you will be able to receive an additional BONUS PAYMENT of £3. To determine whether you will receive the bonus payment, we will randomly select 10 responses from the task you complete. If 8 of these 10 responses are correct, you will receive the bonus payment (in addition to the base-rate pay)."

This incentivisation scheme is created to maximise motivation and engagement during each trial (as they are not able to predict which trials will be evaluated to determine the bonus), whilst being fit for the level of difficulty of the task. (The 34 participants of our separate pilot study achieve 77.2% accuracy, on average, across all instruction variations and without incentivisation.)

*Feedback Questionnaire*

We included a feedback questionnaire in our task to gather participant insights on aspects such as difficulty, clarity of instructions, helpfulness of instructions, and the number of images they had trouble with annotating. For specific questions and detailed information, please refer to Appendix A.1.

*Demographic Questionnaire*

We included a demographic questionnaire in our task, covering age, ethnicity, current employment status, and the highest level of education. Participants are also queried about their past experiences with online-mediated work, specifying domains and tasks if applicable. Additionally, participants are asked to rate (using Likert scales) their prior experience in annotation tasks (see further Appendix A.1.).

**Procedure**

Before commencing the experiment on Gorilla, participants give their informed consent. Subsequently, they are randomly allocated to one of the six conditions described in the "Design" section. They are told that they would be required to categorise 100 consecutive images of building entrances as either "barrier-free" or "not barrier-free" and are given condition-specific instructions regarding the classification instructions and bonus (if in the incentive conditions). Subsequently, participants classify 100 images of entrances (presented in randomised order) as barrier free or not barrier free. At the end of the classification task, participants complete the demographics questionnaire, followed by the feedback questionnaire. Finally, participants are redirected to Prolific and monetarily compensated for their time. As a baseline, participants are paid £4.50 for completing the study, which on average took them 4.68 minutes to complete. Their accuracy on ten randomly selected images was calculated for participants in the "incentive" conditions. These participants were compensated with an additional £3 payment if an 80% accuracy threshold was met.

IV. RESULTS

**Hypothesis Tests**

Our results suggest that task-instruction design and accuracy-based monetary incentives determine the outcome (i.e., accuracy, perceived helpfulness, and perceived difficulty) of data annotation. In short, we postulated that accuracy improves when annotators are provided with rules and rewarding them financially (hypotheses 1a and 1b), associated with an increase in helpfulness and a reduction in perceived difficulty when annotators are working under rules (hypotheses 2a and 2b).

*Accuracy*

Figure 2 provides first evidence that the effects on accuracy are as hypothesised. Annotators working under rules had higher levels of accuracy (85.6%) than annotators working under incomplete rules (76.8%) or standards (74.9%).[2] The differences between annotators working with or without the monetary incentive vary less significantly. Annotators under the incentive condition had a slightly higher accuracy (80.1%) than those without the incentive (78.2%). The design of our experiment also allows us to test the combined effect of rules and incentives, which is shown in the lower panel of Figure 2. Within each group working under the same instruction design, we observe a tendency for higher accuracy among annotators working with a monetary incentive. This difference is significant (at a 95% confidence level) for annotators working under rules. Here, annotators who worked under the incentive condition reached accuracy levels of 86.7%. The impact of monetary incentives was not significant (95% CI) for annotators in the groups of incomplete rules or standards.

---

[2] As in all graphic representations of the results, we choose 95% confidence intervals as error bars, if not indicated differently.

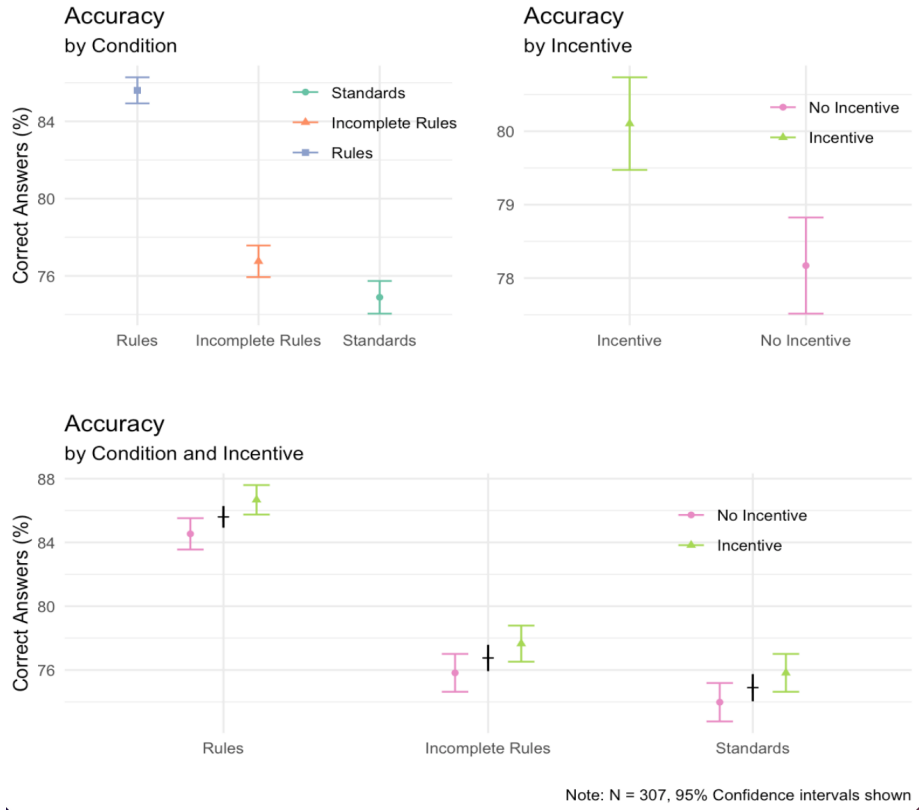

**Figure 2.** Annotators working under rules demonstrate higher accuracy (85.5%) compared to those under incomplete rules (77%) or standards (75%). The combined effect of rules and incentives shows a significant increase in accuracy reaching 87% under the incentive condition.

To account for potentially confounding factors and relate the magnitude of their influence on accuracy to instruction design and monetary incentives, we run a multivariate regression model, shown in Table 1. The regression model has the following structure:

$$\text{Accuracy}_i = \beta_0 + \beta_1 (Condition_{Incomplete})_i + \beta_2 (Condition_{Rules})_i + \beta_3 (Incentive_{yes})_i \\ + \beta_4 (Difficulty)_i + \beta_5 (Helpfulness)_i + \beta_6 (Age)_i \\ + \beta_7 (Uni\ Degree)_i + \beta_8 (Experience)_i + \epsilon_i$$

The model explains the variance in accuracy on an individual level ($y_i$) with features of instruction design, incentives, perceived difficulty (in number of images annotators had trouble with), perceived helpfulness, and individual controls of age, education, and experience.

In comparison to standards, working under rules increases annotators' accuracy by 13 percentage points, as shown in Model 1a of Table 1. Receiving the monetary incentive increases annotators' accuracy by 2 percentage points (Model 1a). In relation to personal characteristics, it is worth noticing that instruction design is roughly 3 times more influential on accuracy than education and roughly 6 times more influential on accuracy than having prior experience in data annotation (Model 1b).

**Table 1.** Rules increase annotators' accuracy by 24 percentage points, the monerary incentive increases annotators' accuracy by 2 percentage points (Model 1a). Rules increase accuracy almost six times more than having experience in data annotation (Model 1b).

|  | Accuracy | | Difficulty | | Helpfulness | |
|---|---|---|---|---|---|---|
|  | Model 1a | Model 1b | Model 2a | Model 2b | Model 3a | Model 3b |
| Condition (Ref: Standards) | | | | | | |
| Incomplete Rules | 0.02** | 0.02* | -1.56 | -1.72 | -0.14 | -0.15 |
|  | (0.01) | (0.01) | (2.19) | (2.18) | (0.29) | (0.29) |
| Rules | 0.13*** | 0.12*** | -4.03* | -4.55** | 0.47** | 0.44** |
|  | (0.01) | (0.01) | (2.19) | (2.19) | (0.28) | (0.29) |
| Incentive (Ref: No Incentive) | | | | | | |
| Incentive | 0.02** | 0.01 | -0.23 | -1.16 | -0.15 | -0.05 |
|  | (0.01) | (0.01) | (1.79) | (1.83) | (0.23) | (0.24) |
| Difficulty | | | | | | |
| number of images (log) |  | -0.01*** |  |  |  | -0.33*** |
|  |  | (0.003) |  |  |  | (0.11) |
| Helpfulness | | | | | | |
| Yes/No |  | 0.01*** |  | 4.43** |  |  |
|  |  | (0.01) |  | (2.09) |  |  |
| Individual | | | | | | |
| Age (years) | 0.0000 | -0.0002 | 0.10 | 0.07 | -0.01 | -0.01 |
|  | (0.001) | (0.001) | (0.09) | (0.09) | (0.01) | (0.01) |
| Uni Degree (0/1) | 0.03** | 0.04*** | -1.80 | -1.53 | -0.10 | -0.09 |
|  | (0.01) | (0.01) | (2.40) | (2.39) | (0.31) | (0.32) |
| Experience (0/1) | 0.02 | 0.02* | -0.58 | -0.25 | -0.03 | -0.05 |
|  | (0.01) | (0.01) | (2.05) | (2.05) | (0.27) | (0.27) |
| Constant | 4.27*** | 4.24*** | 16.83*** | 13.83*** | 0.13 | 0.85 |
|  | (0.02) | (0.02) | (3.95) | (4.17) | (0.51) | (0.58) |
| Observations | 307 | 307 | 307 | 307 | 307 | 307 |
| $R^2$ | 0.31 | 0.34 | 0.02 | 0.03 | 0.08 | 0.10 |
| Adjusted $R^2$ | 0.29 | 0.32 | 0.02 | 0.01 |  |  |
| AIC |  |  |  |  | 430.93 | 420.34 |

Note: *p<0.1, **p<0.05, ***p<0.01

*Perceived Helpfulness and Perceived Difficulty*

Figure 3 suggests that effects on perceived helpfulness and perceived difficulty are as hypothesised in hypotheses 2a and 2b. As regards helpfulness, we found the following distribution of answers on our Likert scale (number of participants in brackets): Hindered a lot (1), Hindered somewhat (13), Neither nor (12), Helped somewhat (139), Helped a lot (142). To account for this skew towards helpfulness in responses, we pool the responses to all points on the Likert scale except for the point "Helped a lot" to create a binary representation of the outcome. Thus, task instructions either "Helped a lot" or did not help a lot. As the column chart on the right in Figure 3 shows, rules had the highest share of annotators responding that the instructions "Helped a lot" (56%, incomplete rules: 40%; standards: 43%). The positive impact of urle-based instructions is similarly confirmed by the multivariate regression model (see Model 3b in Table 1).

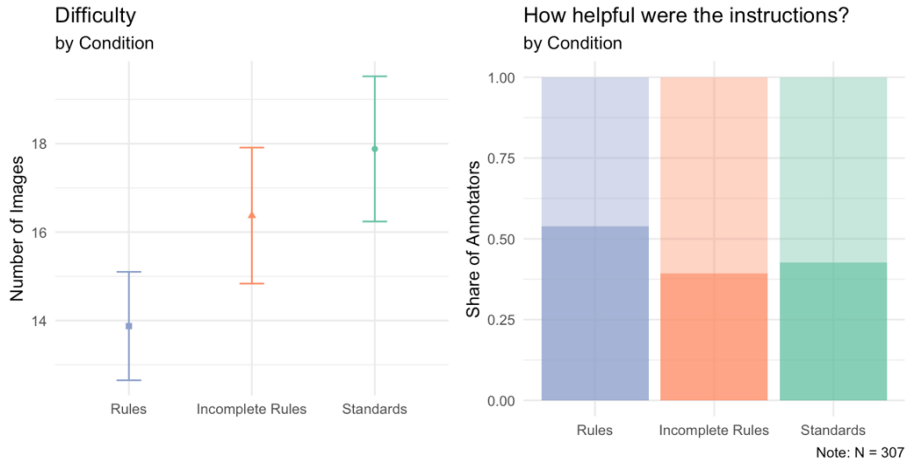

**Figure 3.** Left: Annotators working under rules report having had trouble with fewer images (14) than compared to those working under incomplete rules (16) or standards (18). Right: 56% of annotators working under rules found the instructions to have "Helped a lot", more than those working under incomplete rules (40%) and standards (43%).

## V. Discussion

The purpose of this study was to deliver experimental insights into the working conditions of one type of AI data workers, namely data annotators. Our findings suggest that the performance (accuracy) of annotators at an image labelling task is directly positively affected by choosing rules as task instructions and indirectly by providing an accuracy-based monetary incentive. We hope that our results can help improve work conditions for data annotators around the globe, increase the quality of data sets and thus AI models, as well as contribute to the emerging scholarship on AI data work. In the following discussion, we will revisit our results, lay out the practical and theoretical implications of our experiment, and consider its limitations and directions for future research.

**Practical Implications**

*Organisational Implications*

Our results suggest that AI developers do have an incentive to invest in improving the working conditions of data annotators. By investing in rules and monetary incentives for annotators, they receive higher accuracy in labelling images – especially when both conditions are combined, and annotators are working under rules *and* with a monetary incentive. As the annotators in our study find rules more helpful and report having less trouble classifying images, organisations may also improve worker wellbeing by designing annotation instructions as rules.

Our study did not identify the up-front costs of creating rules. However, an organisation would only have to bear these costs once to then benefit from economies of scale. The higher the number of images labelled, the lower the cost of rules per image. Without including the costs of creating rules, in our study rules were more cost efficient in raising accuracy than the monetary incentive. In our sample, the accuracy for annotators working under standards was 74.9%. Rules improved this accuracy level to 85.6%, or by 14%. It took annotators working under rules 2.95 seconds to complete an image, 0.38 seconds longer than the annotators working under standards needed (2.56 seconds). The best performing annotators were those working under rules and with an additional accuracy incentive of £3, achieving an average accuracy of 86.7%. This equals an increase in accuracy compared to annotators working under standards and without financial incentive of 17%. However, these workers also induced higher costs in two ways.

First, they received a higher payment. In our study, the monetary incentive amounts to two-thirds of the base pay. Experimental testing could determine the optimal (and potentially lower) amount of the incentive payment, hence increasing the efficiency of the monetary incentive. Second, annotators working under rules need, on average, 2.95 seconds per image and thereby significantly longer (15% longer) than those working with standards (2.56 seconds). However, there is no significant difference between the duration of task completion for annotators under rules and for those under incomplete rules. Most likely, this indicates that it is equally time consuming for annotators to check images based on a set of rules, regardless of their completeness, as compared to standards. For the monetary incentive, the picture is clearer. Annotators working with the incentive need 3.14 seconds on average, which is 27% longer than their counterparts without the incentive (2.47). Annotators take much more time per image when a higher financial reward based on performance, in this case accuracy, is offered. Extrinsically motivated annotators presumably work more diligently. If annotators are paid by the hour, the increase in duration per image would of course raise their labour costs. Alternatively, organisations may opt to pay annotators per image or data set, thus neutralising the effect of duration increase on labour costs. That said, annotators in our study labelled a data set of 100 images once in, on average, 4.68 minutes. They thus had little efficiency gains to expect from internalising the rules we gave them for a one-off job. For larger data sets, annotators' expectations may be different, and the time needed per image under the rules condition may approach the time needed under standards. In our study, we indeed saw rules approach standards in task duration as the task progressed. In other words, we found annotators to be 'learning on the job': their performance improved as the task progressed, for both accuracy and duration (see Appendix A.3) .

At this point, the relationship between duration and accuracy warrants revisiting. As mentioned in the "Results" section, receiving the monetary incentive increases annotators' accuracy by 2 percentage points. However, we also find that one percent increase in task duration improves accuracy by 5 percentage points. Does duration thus mediate the effect of incentives? As Annex A.2. shows, accuracy is always higher when the monetary incentive is present, regardless of how long participants took per image (Figure A2, right chart). While accuracy does increase as duration increases, we do not find the slopes of both monetary incentive conditions to be significantly different. Therefore, we do not find duration mediating the effect of the monetary incentive. Instead, our findings suggest that annotators with the monetary incentive may work more diligently, thus taking more time to assess the features of the building entrances and the annotation instructions. For the instructions conditions, the left chart of Figure A2 indicates that irrespectively of duration, workers under rules have significantly higher accuracy than those working with different instructions. In addition, interestingly we see a steeper slope for the rules group indicating that longer task completion is associated with higher accuracy. Taking more time at annotating an image under rules translates into higher accuracy while it does not for the other instruction conditions.

Our findings are furthermore relevant for organisation which are concerned with worker wellbeing. Annotators working under rules found their instruction more helpful and their task less difficult than those working under incomplete rules and standards.

> *Organisations should consider task instruction design and monetary incentives as important factors in the cost-benefit analysis of generating high-quality data sets. These factors are also relevant for human resource management, as instruction design also improves the perceived helpfulness for and the perceived difficulty of annotators' tasks and thus, ultimately, AI data worker wellbeing.*

*Policy Implications*

Our results hold implications for policies on AI data work as well as for the governance of AI systems. First, offering a monetary incentive and, thus, higher wages to annotators improves working conditions in AI data work. Policymakers should thus require adequate compensation for data annotation. The (industry-funded) non-profit organisation Partnership on AI suggests paying data annotators "at least the living wage for their location" (Partnership on AI, n.d.). Considering that much data annotation work is outsourced to regions with comparatively low salaries in the Global South, the additional costs of raising annotators' pay do not seem prohibitively high (and may well be offset by gains in accuracy). The monetary incentive in our study added two-thirds of the base pay. For data annotators working at less than two dollars an hour, such as those reportedly (Perrigo, 2023) labelling texts for OpenAI, this would amount to paying a bit more than an additional dollar per annotator per hour. Going beyond the local living wage should thus not be excessively costly, especially for large tech companies from the Global North. We thus hope that our study results may contribute to greater fairness in outsourcing labour.

Second, improving work conditions in data annotation can make a positive contribution to the accuracy of annotation and, hence, of AI models. Our results reinforce previous findings on the positive effects of increasing information density in annotation instructions on accuracy (Rädsch et al., 2023). The particular set of images in our study—concerning the accessibility of buildings to people with physical disabilities—emphasises the societal importance of well-trained AI systems. Recent research shows how machine-learning models suffer from bias when tasked with predicting the location of curb ramps in cities (Deitz, 2023). For wheelchair riders, curb ramps are essential infrastructure that allows autonomous navigation. In our study, annotators made significantly fewer mistakes in labelling images correctly as barrier-free compared to labelling images correctly as non-barrier-free. There are only minor differences for barrier-free images, ranging from an accuracy of 86% for labelling under standards without incentives up to 95% accuracy for annotators working with rules and incentives. For non-barrier-free images, the difference of norms is accentuated. Here, standards yield an average accuracy of 60% while rules increase accuracy on average to 76%. If left unaddressed, this difference could lead to bias in an AI system trained with such data, with the AI potentially missing important barriers to accessibility. This is particularly noteworthy when extending our results to data annotation in general. We expect that in many annotation tasks the likelihood of annotators committing Type I and Type II errors will be unbalanced—that is, the likelihood of annotating a false positive or a false negative will be different depending on the how the task is structured. Therefore, improving data annotation through experimentally testing annotation instructions can help safeguard ethical values and legal rights such as non-discrimination in AI-driven decision-making.

For an annotation task such as the one in our study, rules appear to be the best choice of instruction designs. However, rules will not be a desirable choice for every data annotation task. As mentioned, rules can be detrimental to pluralism when it is valuable. Moreover, standards allow yielding to annotators' domain expertise. Medical expertise, for example, is highly valuable (but regularly too costly) to obtain for annotating data sets of biomedical images (Freeman et al., 2021; Rädsch et al., 2023).

Finally, our research holds implications for the emerging ethical and regulatory demand of installing and maintaining human oversight of AI (Green, 2022; Laux, 2023; Sterz et al., 2024). Take the European Union's AI Act as an example. Article 14 AI Act requires human oversight of high-risk AI systems, i.e., an effective human involvement in AI systems to mitigate "risks to health, safety or fundamental rights" (Art. 14(2)). Our study suggests that the choice of annotation instructions and annotators' pay influences the quality of a data set on which an AI system is trained. The procedures of data annotation are thus relevant for risk mitigation. Our results amplify the importance of publishing annotation instructions as a precondition for scientific review (Gebru et al., 2018; Maier-Hein et al., 2018; Rädsch et al., 2023) and—in a regulatory context—independent audits and human oversight of AI systems. Human oversight workers should know under which instructions and work conditions an AI training data set has been annotated. It is encouraging to see that the International Organization for Standardization (ISO) requires organisations to provide or demand documentation about acquired data sets, including "details of data labelling and enhancing" in its management system standard on AI (International Organization for Standardization (ISO) & International Electrotechnical Commission (IEC), 2023, p. 38). In the EU, the AI Office of the European Commission could facilitate the collection and distribution of such labelling information (Pouget & Laux, 2023).

Our study further demonstrates the benefits of experimentally testing human labour practices in the AI lifecycle for achieving important goals of AI governance, such as higher-quality AI data sets. It also shows that such testing is feasible at relatively low costs by recruiting participants from online labour platforms. This is encouraging for demands to make experimental testing of AI governance tools (such as human oversight schemes) mandatory before a high-risk AI system is deployed (Green, 2022; Laux, 2023).

> *The improved quality of data sets brought about by offering monetary incentives for data annotators provides an argument for fair wage policies. The choice of instructions for data annotators and their renumeration should be documented. Making such documentation available also for auditors will support human oversight of AI and, hence, the protection of ethical values and fundamental rights.*

**Theoretical Implications**

Our study presents experimental evidence on the combined effects of task instruction design and renumeration policies on the performance of AI data workers. It thus strengthens claims about the important role of human labour and its working conditions for AI production.(Crawford, 2021; Miceli & Posada, 2022; Muldoon, Cant, et al., 2024; Tubaro et al., 2020).

Our findings corroborate several tenets of the law and economics literature on rules versus standards. As predicted by the literature, annotators fare better under rules as the behaviour required by the task is frequent and homogeneous. Rules were also perceived as easier to apply and more helpful (see Table 1/Figure 3), indicating that decision costs are lower for annotators working under rules. As predicted, error costs are higher when annotators work with standards. In our study, annotators need continuously less time per image as they progress in their labelling task. We interpret this result as annotators "learning on the job". Part of that learning could be adaptive behaviour exhibited by annotators which has been theorised by Schauer in his "convergence thesis": those working under standards gradually turn these into rules (Schauer, 2003). Note that this self-made "rule" would already be internalised by the annotators and would thus not inflict the same decision costs on the annotator as working under externally derived rules. However, we tested the performance of decisionmakers under rules versus standards outside of the literature's main case of application, namely that of adjudication and administrative decision-making. In our context of data annotation, the motivations of actors may likely be too different to confirm or reject the literature's theoretical assumptions about adaptive behaviours. Nevertheless, we suggest that data annotation and comparable tasks such as content moderation are valuable domains for examining the effects of rules and standards, as decisions are made more frequently (albeit arguably much less deliberately) than in adjudication or administrative decision-making.

Lastly, our study presents further experimental evidence for the psychological claim that monetary (extrinsic) incentives can boost performance (as measured in accuracy) in a task which workers are presumably not intrinsically motivated to do.

**Limitations and Future Research Directions**

One limitation of our research lies in our recruitment of participants from online labour platforms. In as much as data annotation is done in-house or is outsourced to companies which permanently employ their annotators, treatment effects could be different. Another limitation lies in our monetary incentive to be performance-based. Historically, performance-based pay has been used as a labour-cost cutting strategy by corporations (Acemoglu & Johnson, 2023, p. 258). It may be detrimental to low-performing workers. As mentioned, organisations (or researchers) may determine an optimised pay structure, adjusting the base-pay rate and/or a performance-based bonus.

A potential objection to our methods could be seen in the fact that the rules condition is equivalent to the study's ground truth of what the researchers defined as a barrier-free entrance. We labelled the images according to the same set of rules that was given to participants in the rules condition. Hence, whether an image was correctly or incorrectly labelled by participants was measured by the researchers against the instructions in the rules condition.[3] We would like to offer three responses to this possible objection. First, even though annotators in the rules condition were given the ground-truth criteria for barrier-free entrances, they still made errors in tracking the true conditions for accessibility. Second, as argued above, accessibility norms are one domain in which pluralism, i.e., annotators' different interpretations of what renders an entrance barrier-free, is not desirable. For some domains such as reinforcement learning used in medical AI development, it may be desirable to replace untrained annotators with domain experts (Freeman et al., 2021; Mittelstadt et al., 2023), who may perform better under standards if standards allow them to draw on their expertise to a greater extent than rules would. Third, as already mentioned in the discussion on study design we selected the set of rules for what constitutes a barrier-free entrance by drawing on existing architectural codes for accessibility in buildings in Germany, the location of the vast majority of entrances in the set of 100 images.

Lastly, we only recruited participants who were based in Europe and North America. We thus assume that they have some experience of architecture that is comparable to the European building entrances depicted in our data set. Judging the accessibility of buildings to people with disabilities is presumably influenced by cultural context as well as exposure to architectural designs. Future research could thus repeat the experiment with participants based outside of Europe and Northern America. Comparing the results of this follow-up experiment with the results in our study could shed light on the effects of outsourcing data annotation work to culturally diverse regions (on the influence of cultural cognition on AI training data, see (Atari et al., 2023); on content moderation, see (Jiang et al., 2021)).

## VI. CONCLUSION

The quality of data sets is increasingly being recognised as a crucial factor in training AI models that can responsibly be deployed (Whang et al., 2023). Annotation errors can have a cascading negative effect over the entire life cycle of an AI system (Rädsch et al., 2023; Sambasivan et al., 2021). Our study showed that AI developers can expect a return on accuracy when investing in designing task instructions as clear rules rather than vague standards and providing a performance-based monetary incentive . The implications of our findings go beyond cost efficiency. They affect the work conditions and hence the wellbeing of AI data workers. Moreover, they highlight the importance of empirically investigating human labour in AI production beyond engineering and data analytics. After all, the processes of data annotation impact the quality of an AI system and hence are crucial for mitigating risks to safety and fundamental values that AI may create.

---

[3] We tested agreement on annotating the image set amongst two researchers (n=20 images) and report an agreement score of 100%.

# APPENDIX

## A.1. Participant Survey

**Table A1:** Means and standard deviations for the surveys on participant demographics and prior experience

| Variable | N | Mean | SD | Min | Max |
|---|---|---|---|---|---|
| <u>Participants</u> | | | | | |
| Age | 307 | 30 | 9,7 | 20 | 66 |
| Gender | | | | | |
| ... *Female* | 135 | 0,44 | | | |
| ... *Male* | 168 | 0,55 | | | |
| ... *Non-binary* | 4 | 0,01 | | | |
| Qualification | | | | | |
| ... *No Uni* | 51 | 0,17 | | | |
| ... *Uni* | 256 | 0,83 | | | |
| Experience | | | | | |
| ... *No* | 227 | 0,74 | | | |
| ... *Yes* | 80 | 0,26 | | | |
| <u>Treatments</u> | | | | | |
| Condition | | | | | |
| ... *Standards* | 101 | 0,33 | | | |
| ... *Incomplete Rules* | 102 | 0,33 | | | |
| ... *Rules* | 104 | 0,34 | | | |
| Incentive | | | | | |
| ... *No Incentive* | 153 | 0,5 | | | |
| ... *Incentive* | 154 | 0,5 | | | |
| <u>Outcomes</u> | | | | | |
| Accuracy (% of correct answers) | 307 | 79 | 8,3 | 41 | 99 |
| Duration (in seconds) | 307 | 2,8 | 1,7 | 0,9 | 16 |
| <u>Feedback (self-reported)</u> | | | | | |
| Clarity | | | | | |
| ... *Clear* | 168 | 0,55 | | | |
| ... *Neither clear nor unclear* | 21 | 0,07 | | | |
| ... *Unclear* | 20 | 0,07 | | | |
| ... *Very clear* | 97 | 0,32 | | | |
| ... *Very unclear* | 1 | 0 | | | |
| Difficulty | | | | | |
| ... *Not Difficult* | 272 | 0,89 | | | |
| ... *Difficult* | 35 | 0,11 | | | |
| Helpfulness | | | | | |
| ... *Helped a lot* | 139 | 0,45 | | | |
| ... *Helped somewhat* | 142 | 0,46 | | | |
| ... *Hindered a lot* | 1 | 0 | | | |
| ... *Hindered somewhat* | 13 | 0,04 | | | |
| ... *Neither helped nor hindered* | 12 | 0,04 | | | |
| Trouble | | | | | |
| ... *Neither helped nor hindered* | 307 | 16 | 16 | 0 | 100 |

## A.2. Duration

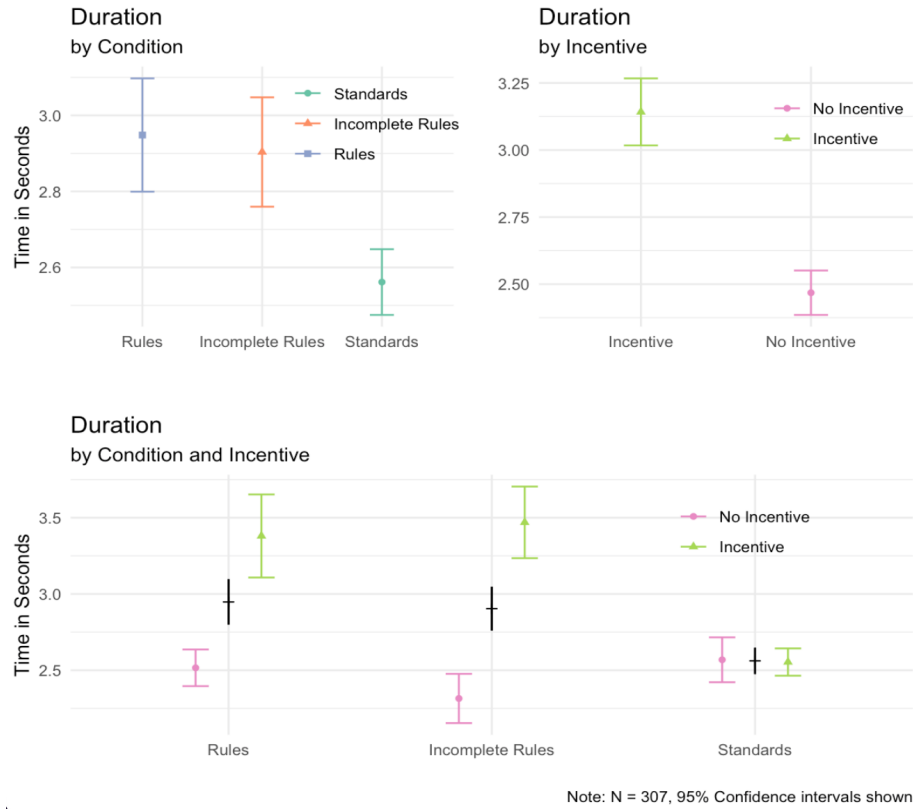

**Figure A1.** Working under rules requires significantly more time per image (2.95 seconds) than working under standards (2.56 seconds). Annotators working with incentives spend on average 27% more time per image compared to their counterparts working without additional financial reward.

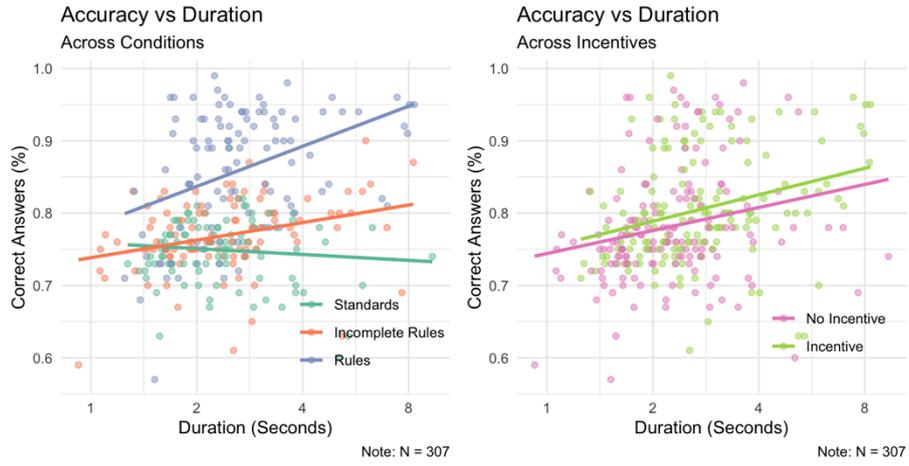

**Figure A2.** On average, duration and accuracy have a positive relationship. On the left-hand side we see that this positive association is even amplified under the rules condition. On the right-hand side we see no significant differences of incentives on the relationship between duration and accuracy but for a higher base level of accuracy for monetary incentives.

## A.3. Learning on the Job

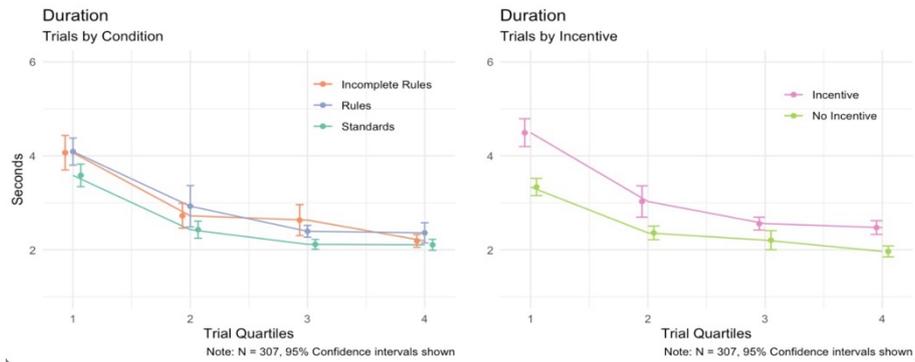

**Figure A3:** Learning measured as change in duration over time.

As Figure A3 indicates, task duration declines as the number of images labelled increases. Annotators working with standards showed a consistently shorter task duration than annotators working with rules and incomplete rules, as do annotators with no monetary incentive.

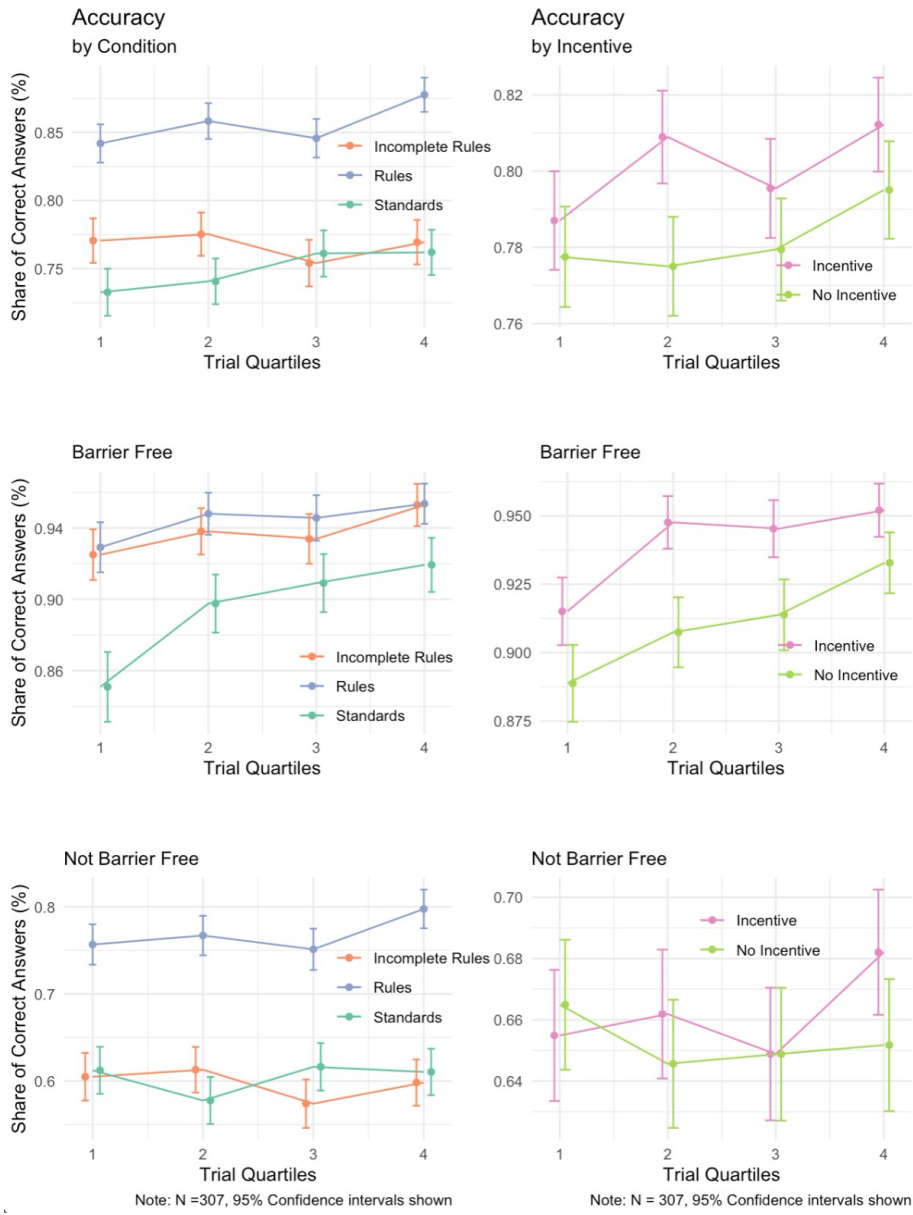

**Figure A4:** Learning measured as change in accuracy over time.

As Figure A4 indicates, annotators' accuracy tends to increase as the number of images labelled increases. However, the change in accuracy is mainly driven by an increase in accuracy for barrier-free images (Figure A4 middle left and right).

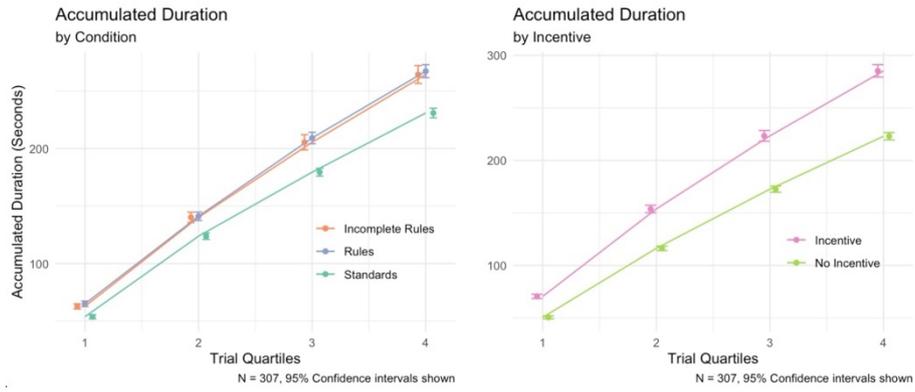

**Figure A5.** The longer the annotation lasts, the more the accumulated time spent on the tasks diverges between the groups. While duration gaps were marginal at the beginning, towards the end of the exercise, annotators under rules needed significantly longer, as did annotators with incentives.

## A.4. Image Type

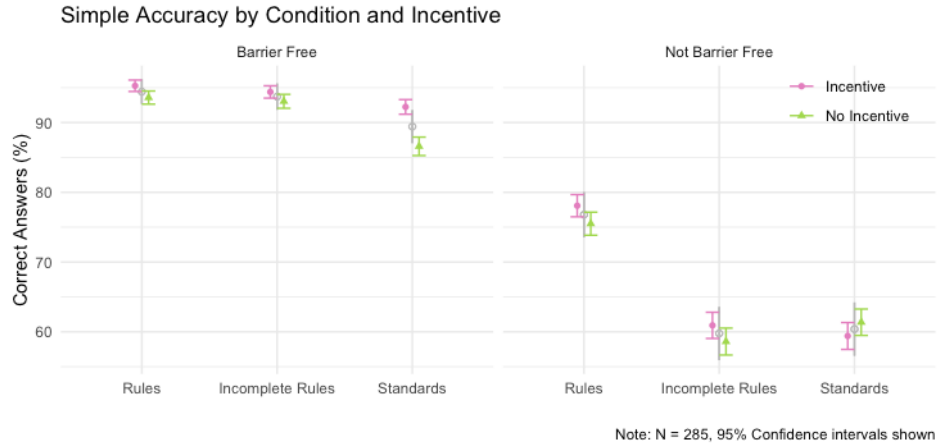

**Figure A6.** For barrier-free images, accuracy ranges from 86% for labelling under standards without incentives to 95% accuracy for annotators working with rules and incentives. For non-barrier-free images, accuracy is lower across all conditions, and the difference of instructions is accentuated.

We note that accuracy is on average higher when annotators are asked to label barrier-free images, across all conditions, as shown in Figure A5. This suggests that barrier-free images are easier to annotate, regardless of which instruction design has been chosen and whether an incentive has been offered or not. The effects of norms and incentives play out differently, therefore, depending on the type of image. There are only minor differences for barrier-free images, ranging from an accuracy of 86% for labelling under standards without incentives up to 95% accuracy for annotators working with rules and incentives. For non-barrier-free images, the difference of norms is accentuated. Here, standards yield an average accuracy of 60% while rules increase accuracy to 76%, on average. Annotators under standards are not performing better than those under incomplete rules. This potentially indicates that standards are less helpful for non-barrier-free images, as incomplete rules should lead to low levels of accuracy, since a crucial rule for identifying barriers is missing in the description. As shown above (see Materials section), the rules condition mentioned the need for a handrail for an entrance to be judged barrier-free which was not mentioned in the incomplete rules condition. Columns three and four in Table 1 confirm these observations. The effect of rules on accuracy is fifteen times higher for non-barrier-free images than for images showing a barrier-free entrance. Accordingly, we assume that the effect of norm design is more pronounced for images which are more challenging to annotate.